\documentclass[useAMS,usenatbib]{mn2e}
\usepackage{graphicx}
\title[Lithium Abundance Anomalies in Stars with Planets]{Parent Stars of Extrasolar Planets - XIII. Additional Evidence for Li Abundance Anomalies}
\author[G.\ Gonzalez]{G.\ Gonzalez$^{1}$\\
$^{1}$Department of Physics and Astronomy, Ball State University, Muncie, IN 47306 USA\\
}

\begin{document}

\date{Accepted ??. Received ??; in original form ??}

\pagerange{\pageref{firstpage}--\pageref{lastpage}} \pubyear{??}

\maketitle

\label{firstpage}

\begin{abstract}
We report the results of our analysis of new high resolution spectra of 37 late-F to early-G dwarf stars for the purpose of deriving their Li abundances. Most of the stars were selected from the large Valenti and Fischer compilation and had unknown Li abundances prior to the present study. When the new data are combined with data from our previous studies on this topic and analyzed in a similar way, we find, again, that stars with planets near the solar temperature are deficient in Li relative to a comparison set of stars. A similar result is obtained when we combine our data with a large database of stellar Li abundances from the literature.
\end{abstract}

\section{Introduction}

In this study we again revisit the question of a possible correlation between the presence of Doppler-detected planets and stellar Li abundance. Several studies \citep{is04,tk05,gg08,is09,gg10,del14} indicate that stars with planets (SWPs) have lower Li abundances compared to stars without detected planets over a limited range in effective temperature (T$_{\rm eff}$) near the solar value. However, other studies \citep{lh06,bau10,gh10,ram12} have failed to confirm this pattern. Therefore, despite having received attention for about a decade from several inpedendent groups, this question is still unsettled.

This question is important, because the Li abundance in a star's atmosphere is sensitive to a number of processes. These include gradual destruction of Li by canonical convective mixing in a star's envelope \citep{pin97}, enhanced destruction of Li from rotationally-induced mixing \citep{pin90}, and increase \citep{lg03,ash05} or decrease \citep{tv12} of surface Li abundance from accretion of planetary material. Rotational mixing (and the associated destruction of Li) can be enhanced by external torques, such as from the presence of stellar companions \citep{rd95}, a migrating planet \citep{cas09}, or a protoplanetary disk \citep{bou08}. It is this last mechanism that is most relevant to the present study. The formation of Doppler-detectable planets is more likely if a star is accompanied by a long-lived protoplanetary disk, and the disk is likely to slow the rotation of the star \citep{matt12}. \citet{gg11} showed that SWPs tend to have smaller vsini values than non-SWPs, confirming the link between planet formation and stellar rotation.

The question of possible differences in Li abundance between SWPs and non-SWPs is therefore an important part of this puzzle and is an important test of the protoplanetary disk-stellar spin-down model. However, disentangling the various factors that influence Li abundance is a difficult task that requires large sample sizes.

In \citet{gg10}, we compared the Li abundances of 50 SWPs and 49 comparison stars; the number of stars we actually employed in the analysis was less than this, as stars having only upper limits on Li abundance were not used in the comparison. In the present study, we seek to improve on that study by increasing the number of Sun-like comparison stars with Li abundance determinations. Our primary list for selecting targets is \citep{vf05}. The stars in that study have been searched for planets with the Doppler method, and the stars also have accurately known atmospheric parameters. However, most of them are still lacking Li abundance determinations; the spectra used by Fischer and Valenti did not include the Li feature at 6707 \AA.

The purpose of the present study is to test again the claim that the Li abundances of Sun-like SWPs are different than those of similar stars without known planets. In Section 2 we describe our new spectroscopic observations and Li abudance analyses. In Section 3 we compare SWPs and stars without detected planets. We present our conclusions in Section 4.

\section{Observations and analyses}

We observed 32 stars from the \citep{vf05} study; three stars (HD 13931, 30562, 38858) are SWPs, and the remainder have no detected planets. The values of T$_{\rm eff}$ listed by \citep{vf05} for these stars range from 5711 to 6326 K, and the average is near 5850 K. In addition, we observed the solar analogs HD 88084 and 208704, which are listed by \citet{sou10} as comparison stars in their study of Li abundances in SWPs and comparison stars; however, they only have upper limits on the Li abundances of these two stars. We also observed the SWPs HD 13908, 60532, 197037, and 220689, only one of which (60532) had been included in our previous studies.

We obtained spectra of 36 of our 37 target stars on December 13-16, 2013 using the McDonald Observatory 2.1-m Otto Struve telescope and Sandiford spectrograph, which is a Cassegrain echelle design \citep{mcc93}. The spectrograph was set to cover the wavelength range 5450-6800 \AA. Two or three spectra of each star were obtained, and exposure times were adjusted to give a similar net S/N ratio for each star. A solar spectrum was obtained via reflected light off the Galilean Moon Callisto, and the hot star Regulus was observed. The resolving power of the spectra is about 53,000, and the S/N ratio at 6700 \AA\ is in the range 300-350 per pixel.

In addition, spectra of HD 52711 were obtained for us by Kyle A. McCarthy with the McDonald Observatory 2.7-m telescope and 2dcoud\'e spectrograph during our run; the spectrograph was setup to cover a very similar wavelength range to what we had used in our previous studies of SWPs (the spectra also have the same resolving power). The S/N ratio of the combined spectrum of HD 52711 is about 700 per pixel near 6700 \AA.

The spectra were reduced with the standard software tools available in IRAF \footnote{IRAF is distributed by National Optical Astronomy Observatories, operated by the Association of Universities for Research in Astronomy, Inc., under contract with the National Science Foundation.}following the same procedures described in our previous series of studies on SWPs, except for one change. For the Sandiford echelle spectra we divided out telluric absorption lines with the IRAF TELLURIC task using our high S/N ratio spectra of Regulus.

\subsection{Measurement of equivalent widths}

For the measurement of Fe line equivalent widths (EWs), we again employed the software package DAOSPEC, which we had previously used in \citet{gg10} (note, for HD 52711 we employ the same settings for DAOSPEC that we used in that prior study). For the Sandiford spectra, we setup the parameters in DAOSPEC to analyze each spectrum over the wavelength interval 5490 to 6780 \AA, and we set the continuum fitting Legendre polynomial order to 1 (this is adequate, since the spectra had already been continuum-normalized in a previous step with IRAF). DAOSPEC typically identified 2000 lines in each spectrum with EWs values $\ge 3$ m\AA.

DAOSPEC also requires an input linelist. For the Sandiford spectra, our input linelist consists of 59 Fe I and 8 Fe II lines. DAOSPEC matches the lines in the input linelist with lines it has identified in a given spectrum. In order to pass to the next step in the analysis, the EW value determined by DAOSPEC for each line in our linelist must satisfy the following additional criteria. It must have an EW value $\ge 3$ m\AA\ and $\le 120$ m\AA, and the ''quality parameter'' value $\le 1.6$ (this parameter compares the residuals in the immediate neighborhood of a line to the overall residuals).

\subsection{Stellar atmospheric parameters}

As with our previous studies in this series, we use the program MOOG \citep{s73} with the model atmospheres of \citet{k93} to derive stellar atmospheric parameters and chemical abundances. For the present study we employed the 2002 version of MOOG.\footnote{Source code of the most recent version is available at http://www.as.utexas.edu/~chris/moog.html} We have calculated a new set of solar-based $gf$-values using the following procedure (very similar to the procedure used in \citet{gg10}).

First, we determined solar EWs with DAOSPEC from the spectrum of Callisto obtained during our run. Then, we selected 21 Fe I lines (between 5490 and 6800 \AA) with high quality EW values that appear in Table 1 of \citet{gs99}, which lists high-quality laboratory $gf$-values for Fe I lines. Next, we determined the solar abundance of Fe from each of these Fe I lines using MOOG and adjusted the microturbulence velocity parameter, $\upsilon_{\rm t}$, to minimize the dispersion; we found a best fit $\upsilon_{\rm t}$ value of 1.2 km~s$^{\rm -1}$. Then, using this value of $\upsilon_{\rm t}$, we adjusted the $gf$-values such that all the Fe I lines used in this work (59) gave an abundance A(Fe) $= 7.470$\footnote{A(Fe) $= \log$ (N$_{\rm Fe}$/N$_{\rm H}) + 12$}; we determined the 8 Fe II line $gf$-values using these same parameter values.

We calculated the stellar atmospheric parameters and their uncertainties using the same procedures we used in our previous papers (see \citet{gg10} and papers cited therein for details on the methods). In brief, LTE is assumed and the stellar parameters are determined assuming excitation and ionisation equilibria. T$_{\rm eff}$ is determined from Fe I lines by requiring that their abundances no not display a trend with their lower excitation potentials, and surface gravity is determined by requiring that the mean Fe I and Fe II abundances be equal (i.e., ionization equilibrium). The microturbulence velocity parameter is determined by requiring that the Fe I abundances do not display a trend with the reduced EW values. In practice, a solution is reached when all three of these criteria are simultaneously met. Errors are propagated statistical uncertainties.

We list the results of our Fe line analysis in Table 1. Note, in deriving the stellar parameters for each star we iterated to the final solution. In the first round, we checked manually for obviously discrepant (more than about 3$\sigma$) Fe line abundances and removed them prior to the second (final) round. Typically, 45-55 Fe I lines and 6-8 Fe II lines were retained for each star. For the case of HD 52711, we employed the linelists and $gf$-values that we used in \citet{gg10}.

Comparing our results to those of \citet{vf05} for the 32 stars in common between our studies, we find that $\Delta$T$_{\rm eff} = 9 \pm 48$ K and $\Delta$log g $= 0.02 \pm 0.10$ dex (in the sense of our values minus theirs). Both offsets are very small, and in each case the scatter of the differences is consistent with the uncertainties quoted in the studies.

\subsection{Lithium abundances}

We determined the Li abundance for each star using spectrum synthesis with MOOG. We employed the same methods described in \citep{gg10}; our adopted solar Li abundance is A(Li) = 0.96.

We estimate the uncertainty of the Li abundance due to noise and unmodelled lines in a typical observed spectrum to be about $\pm$ 0.06 dex (the uncertainty for HD 52711 is slightly smaller, given its higher S/N ratio). The total uncertainty of the Li abundance is based on the quadrature sum of this estimate and the uncertainty of Li due to the uncertainty of T$_{\rm eff}$. The results of our spectroscopic analyses are listed in Table 1. For 12 of the stars in our program we could only determine upper limits on the Li abundances. We will present our abundance results for elements other than Li and Fe in a separate paper.

There is one star in common between the present work and \citet{gg10}, HD 60532. The stellar parameters determined in the two studies are consistent at the level of about $1.5\sigma$; the Li abundance values are nearly the same. For the purpose of the analysis presented below in Section 3, we will use a weighted average set of parameters for this star.

Ten stars from the present work are included in \citep{ram12}, which is a compilation of Li abundances for 1381 FGK dwarf and subgiant stars determined by them and also drawn from the literature. Four of these 10 stars have only upper limits on the Li abundance in either their tabulation or ours. The average difference in Li abundance for the remaining six stars between the two studies is $-0.14 \pm 0.09$ dex (ours minus theirs); this difference is likely due to the different  adopted solar abundance values. The average $\Delta$T$_{\rm eff}$ for the 10 stars in common is only $8 \pm 42$ K.

We show the Li abundances determined in the present work in Figure 1. In the analysis presented below in Section 3, we will only make use of the stars with Li detections.

\subsection{Derived stellar parameters}

We determined age, mass and $\log g$ values listed in Table 1 for each star from stellar isochrones. We employed our T$_{\rm eff}$ and [Fe/H] values with M$_{\rm v}$ calculated from the new reduction of the {\it Hipparcos} parallaxes \citep{van07} with a Bayesian parameter estimation method \citep{da06}.\footnote{We used Leo Girardi's web program PARAM v1.3 to calculate these quantities. Since this is a newer version of the software than we had used in \citet{gg10}, we also  calculated new derived stellar parameters for all the stars in that study. See: http://stev.oapd.inaf.it/cgi-bin/param\_1.3} In calculating M$_{\rm v}$ for each star, we used the {\it Hipparcos} magnitudes corrected according to the prescription in Table 1 of \citep{bes00} to convert them to the Johnson system. The mean difference between our spectroscopic log g values and the parallax-derived (photometric) values is 0.07 $\pm$ 0.08 dex. Thus, within the quoted error, the photometric log g values are consistent with the spectroscopic values.

\begin{table*}
\centering
\begin{minipage}{160mm}
\caption{Parameters of the program stars determined from our spectroscopic analyses. Derived parameters based on stellar isochrones are given in columns 8 to 10.}
\label{xmm}
\begin{tabular}{rrcccccccc}
\hline
Star & & T$_{\rm eff}$ & log g & $\zeta_{\rm t}$ & [Fe/H] & log $\epsilon$(Li) & mass (M$_{\odot}$) & log g & age (Gyr)\\
HD & HIP & (K) & & (km~s$^{\rm -1}$) & & & & & \\
\hline
5372 & 4393 & $5855\pm40$ & $4.40\pm0.06$ & $1.35\pm0.08$ & $~~0.18\pm0.03$ & $<0.93\pm0.10$ & $1.10\pm0.02$ & $4.38\pm0.03$ & $2.6\pm1.5$\\
10086 & 7734 & $5709\pm46$ & $4.43\pm0.05$ & $1.37\pm0.10$ & $~~0.08\pm0.03$ & $~~~1.55\pm0.07$ & $1.02\pm0.03$ & $4.46\pm0.03$ & $2.2\pm1.8$\\
13043 & 9911 & $5862\pm39$ & $4.25\pm0.06$ & $1.60\pm0.12$ & $~~0.03\pm0.03$ & $~~~1.81\pm0.07$ & $1.06\pm0.02$ & $4.17\pm0.03$ & $7.3\pm0.7$\\
13825 & 10505 & $5738\pm34$ & $4.47\pm0.06$ & $1.18\pm0.06$ & $~~0.21\pm0.02$ & $<0.83\pm0.10$ & $1.06\pm0.02$ & $4.38\pm0.02$ & $4.1\pm1.3$\\
13908 & 10743 & $6164\pm57$ & $4.00\pm0.08$ & $2.00\pm0.20$ & $-0.06\pm0.04$ & $<1.18\pm0.08$ & $1.24\pm0.05$ & $4.04\pm0.04$ & $3.9\pm0.6$\\
13931 & 10626 & $5828\pm39$ & $4.30\pm0.05$ & $1.50\pm0.10$ & $~~0.00\pm0.03$ & $~~~1.50\pm0.07$ & $1.01\pm0.02$ & $4.28\pm0.03$ & $7.4\pm1.1$\\
16548 & 12350 & $5713\pm38$ & $4.01\pm0.05$ & $1.35\pm0.06$ & $~~0.17\pm0.03$ & $~~~2.29\pm0.07$ & $1.20\pm0.03$ & $3.95\pm0.03$ & $5.5\pm0.4$\\
28676 & 21158 & $5886\pm33$ & $4.25\pm0.06$ & $1.44\pm0.11$ & $~~0.06\pm0.03$ & $~~~1.89\pm0.07$ & $1.07\pm0.02$ & $4.23\pm0.03$ & $6.5\pm0.7$\\
30562 & 22336 & $5914\pm46$ & $4.26\pm0.09$ & $1.52\pm0.09$ & $~~0.21\pm0.03$ & $~~~2.55\pm0.07$ & $1.23\pm0.04$ & $4.10\pm0.03$ & $4.5\pm0.7$\\
31253 & 22826 & $6045\pm54$ & $4.10\pm0.06$ & $1.65\pm0.13$ & $~~0.09\pm0.04$ & $~~~1.48\pm0.08$ & $1.19\pm0.04$ & $4.17\pm0.04$ & $4.3\pm0.7$\\
31966 & 23286 & $5727\pm25$ & $4.12\pm0.04$ & $1.35\pm0.05$ & $~~0.08\pm0.02$ & $~~~1.41\pm0.07$ & $1.07\pm0.02$ & $4.07\pm0.03$ & $8.1\pm0.4$\\
32963 & 23884 & $5751\pm33$ & $4.41\pm0.05$ & $1.32\pm0.07$ & $~~0.07\pm0.02$ & $<0.76\pm0.07$ & $1.02\pm0.02$ & $4.42\pm0.03$ & $3.4\pm2.0$\\
36108 & 25616 & $5886\pm45$ & $4.33\pm0.07$ & $2.07\pm0.24$ & $-0.25\pm0.04$ & $~~~1.79\pm0.07$ & $0.95\pm0.02$ & $4.12\pm0.03$ & $9.6\pm0.7$\\
38858 & 27435 & $5777\pm45$ & $4.61\pm0.10$ & $1.20\pm0.14$ & $-0.18\pm0.03$ & $~~~1.34\pm0.08$ & $0.94\pm0.03$ & $4.49\pm0.03$ & $2.7\pm2.1$\\
39881 & 28066 & $5737\pm39$ & $4.34\pm0.06$ & $1.35\pm0.11$ & $-0.13\pm0.03$ & $<0.40\pm0.08$ & $0.93\pm0.02$ & $4.26\pm0.02$ & $10.6\pm0.8$\\
44420 & 30243 & $5781\pm46$ & $4.31\pm0.04$ & $1.34\pm0.09$ & $~~0.25\pm0.03$ & $<0.73\pm0.08$ & $1.09\pm0.02$ & $4.39\pm0.03$ & $2.6\pm1.7$\\
44821 & 30344 & $5761\pm47$ & $4.57\pm0.08$ & $1.40\pm0.10$ & $~~0.08\pm0.04$ & $~~~1.77\pm0.07$ & $1.02\pm0.02$ & $4.49\pm0.02$ & $1.0\pm0.9$\\
44985 & 30552 & $6004\pm41$ & $4.47\pm0.07$ & $1.69\pm0.14$ & $-0.07\pm0.03$ & $~~~1.87\pm0.07$ & $1.05\pm0.03$ & $4.39\pm0.03$ & $2.8\pm1.5$\\
47157 & 31655 & $5734\pm34$ & $4.43\pm0.04$ & $1.12\pm0.06$ & $~~0.34\pm0.02$ & $<0.81\pm0.07$ & $1.10\pm0.02$ & $4.42\pm0.03$ & $1.4\pm1.1$\\
48682 & 32480 & $6170\pm34$ & $4.54\pm0.03$ & $1.65\pm0.08$ & $~~0.13\pm0.03$ & $~~~2.62\pm0.07$ & $1.20\pm0.02$ & $4.36\pm0.02$ & $0.9\pm0.6$\\
50692 & 33277 & $5950\pm26$ & $4.51\pm0.06$ & $1.63\pm0.09$ & $-0.16\pm0.02$ & $~~~1.74\pm0.06$ & $0.99\pm0.02$ & $4.38\pm0.02$ & $5.2\pm0.9$\\
52711 & 34017 & $5885\pm27$ & $4.31\pm0.03$ & $1.20\pm0.11$ & $-0.09\pm0.02$ & $~~~1.80\pm0.05$ & $0.99\pm0.02$ & $4.34\pm0.02$ & $6.4\pm0.9$\\
56303 & 35209 & $5993\pm32$ & $4.45\pm0.06$ & $1.40\pm0.08$ & $~~0.16\pm0.02$ & $~~~2.40\pm0.06$ & $1.15\pm0.02$ & $4.34\pm0.03$ & $2.6\pm1.0$\\
60532 & 36795 & $6230\pm68$ & $3.91\pm0.08$ & $1.90\pm0.26$ & $-0.10\pm0.05$ & $~~~1.66\pm0.10$ & $1.49\pm0.03$ & $3.79\pm0.02$ & $2.5\pm0.2$\\
71881 & 41844 & $5839\pm45$ & $4.27\pm0.06$ & $1.49\pm0.14$ & $-0.05\pm0.03$ & $~~~1.46\pm0.07$ & $0.99\pm0.02$ & $4.26\pm0.03$ & $8.0\pm1.0$\\
76909 & 44137 & $5664\pm40$ & $4.24\pm0.07$ & $1.26\pm0.08$ & $~~0.30\pm0.03$ & $<0.74\pm0.07$ & $1.08\pm0.03$ & $4.21\pm0.04$ & $7.0\pm1.1$\\
86264 & 48780 & $6231\pm74$ & $4.19\pm0.08$ & $2.15\pm0.05$ & $~~0.10\pm0.02$ & $<0.87\pm0.10$ & $1.35\pm0.03$ & $4.08\pm0.04$ & $2.8\pm0.3$\\
88084 & 49728 & $5743\pm51$ & $4.37\pm0.06$ & $1.27\pm0.12$ & $-0.08\pm0.04$ & $~~~0.84\pm0.08$ & $0.95\pm0.03$ & $4.40\pm0.04$ & $6.2\pm2.8$\\
111398 & 62536 & $5752\pm33$ & $4.37\pm0.05$ & $1.40\pm0.11$ & $~~0.08\pm0.02$ & $<0.91\pm0.07$ & $1.03\pm0.02$ & $4.19\pm0.02$ & $8.4\pm0.6$\\
193664 & 100017 & $5866\pm45$ & $4.48\pm0.08$ & $1.70\pm0.15$ & $-0.16\pm0.03$ & $~~~2.01\pm0.07$ & $0.96\pm0.03$ & $4.39\pm0.03$ & $6.0\pm1.8$\\
197037 & 101948 & $6210\pm60$ & $4.54\pm0.08$ & $1.97\pm0.26$ & $-0.17\pm0.04$ & $~~~2.50\pm0.07$ & $1.08\pm0.03$ & $4.37\pm0.03$ & $2.1\pm1.4$\\
208704 & 108468 & $5816\pm45$ & $4.37\pm0.05$ & $1.30\pm0.13$ & $-0.08\pm0.03$ & $<1.03\pm0.08$ & $0.97\pm0.02$ & $4.32\pm0.03$ & $7.9\pm1.5$\\
218133 & 114028 & $5995\pm38$ & $4.46\pm0.07$ & $1.70\pm0.10$ & $-0.06\pm0.03$ & $~~~2.02\pm0.07$ & $1.05\pm0.02$ & $4.31\pm0.03$ & $5.1\pm1.0$\\
218730 & 114424 & $5953\pm39$ & $4.48\pm0.06$ & $1.38\pm0.08$ & $~~0.11\pm0.03$ & $~~~2.38\pm0.07$ & $1.11\pm0.02$ & $4.42\pm0.02$ & $1.0\pm0.9$\\
220689 & 115662 & $5929\pm47$ & $4.36\pm0.06$ & $1.28\pm0.11$ & $-0.01\pm0.03$ & $~~~1.67\pm0.07$ & $1.05\pm0.03$ & $4.37\pm0.04$ & $3.7\pm1.9$\\
221830 & 116421 & $5802\pm52$ & $4.34\pm0.05$ & $1.50\pm0.21$ & $-0.35\pm0.04$ & $<1.02\pm0.08$ & $0.90\pm0.01$ & $4.19\pm0.02$ & $11.2\pm0.4$\\
223238 & 117367 & $5865\pm32$ & $4.40\pm0.05$ & $1.49\pm0.09$ & $~~0.02\pm0.03$ & $~~~1.11\pm0.08$ & $1.04\pm0.02$ & $4.27\pm0.04$ & $6.7\pm0.9$\\
\hline
\end{tabular}
\end{minipage}
\end{table*}

\begin{figure}
  \includegraphics[width=3in]{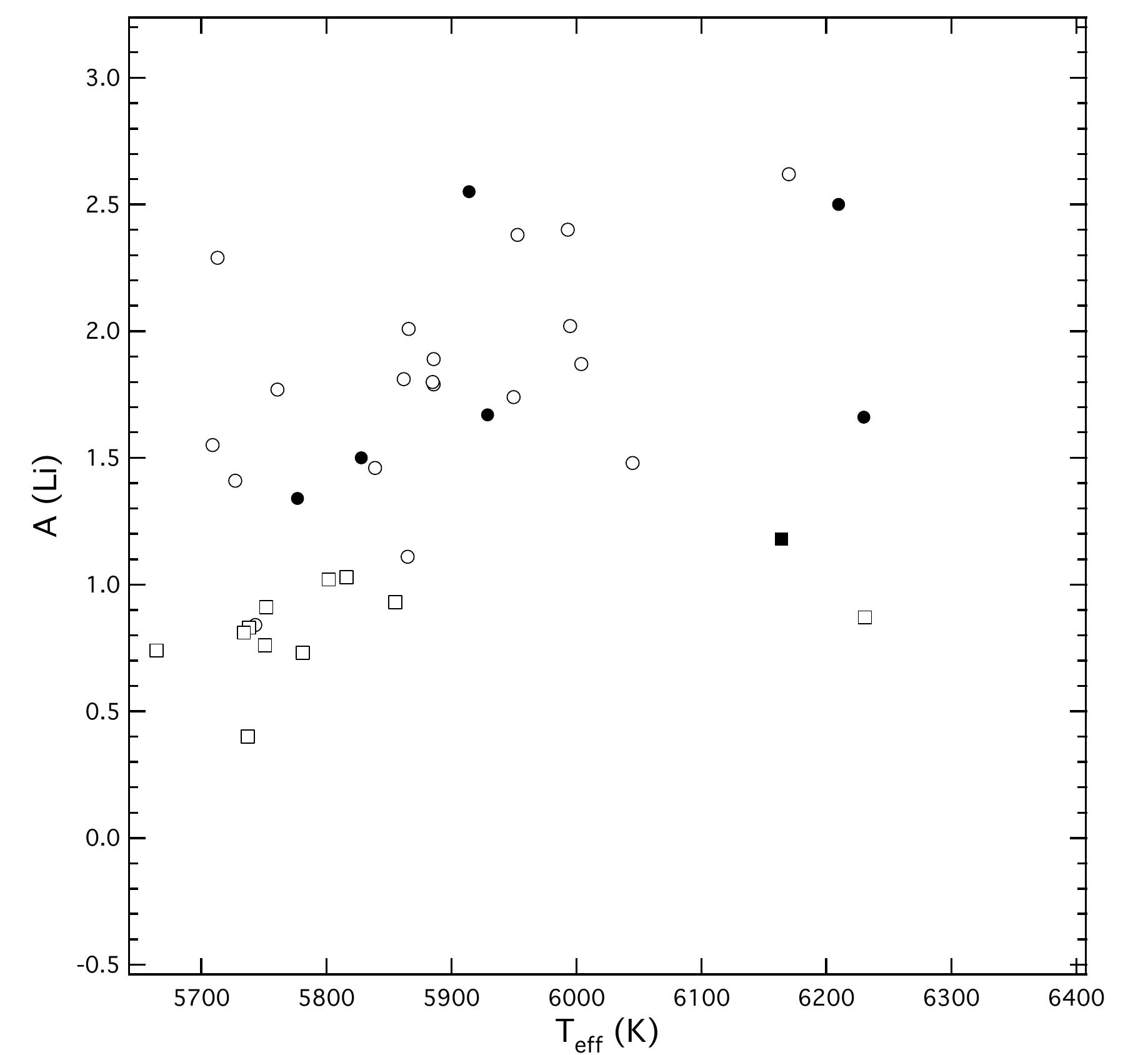}
 \caption{Li abundances versus T$_{\rm eff}$ for SWPs (dots) and stars without planets (open circles) for the stars analyzed in the present work.  Upper limits on the Li abundance are shown for SWPs (filled squares) and stars without planets (open squares).}
\end{figure}

\begin{figure}
  \includegraphics[width=3in]{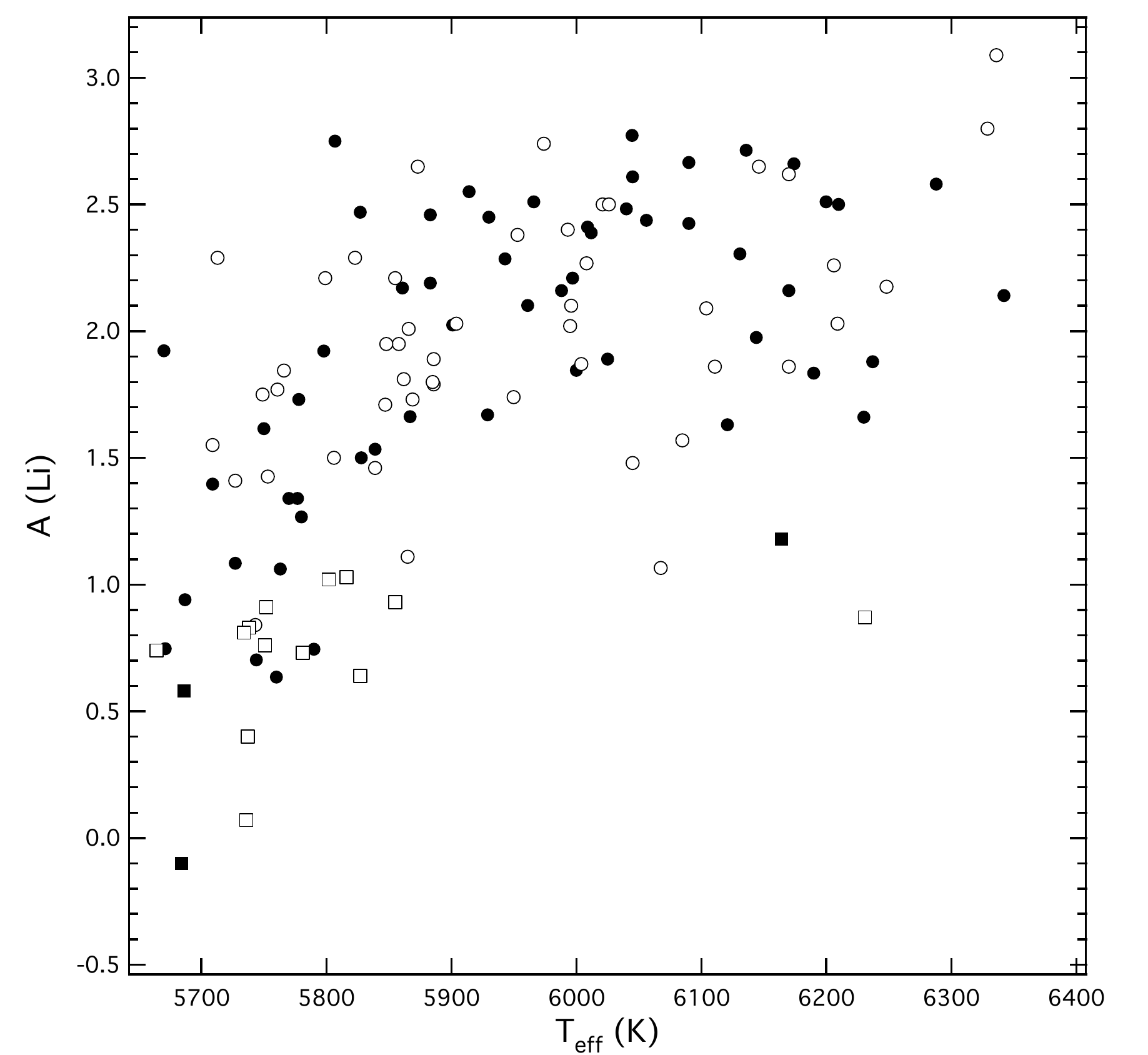}
 \caption{Same as Figure 1 but nowalso  including the stars from \citet{gg10} that fall in the T$_{\rm eff}$ range shown in the figure.}
\end{figure}

\section{Comparison of samples}
\subsection{The new data}

We formed our SWP and comparison stars samples by combining the new results in the present work and the results from \citet{gg10}. We limit our comparison of Li abundances between SWPs and stars without planets to T$_{\rm eff} =$ 5650 to 6350 K. Also, stars having only upper limits for the Li abundances are not included in the analysis. These selection criteria leave us with 50 SWPs and 49 comparison stars from \citet{gg10} and 5 SWPs and 19 comparison stars from the present work.

In \citet{gg08} we introduced a new index, $\Delta_1$, which is a measure of the distance between two stars in T$_{\rm eff}$-[Fe/H]-$\log g$-M$_{\rm v}$ space. We calculated a weighted-average Li abundance difference between a given SWP and all the comparison stars using $(\Delta_1)^{-2}$ as the weight. We also employed this method in \citep{gg10}. We applied this same method to the present dataset. The weighted Li abundance differences are shown in Figure 3. It looks very similar to the equivalent figure (Figure 6) in \citet{gg10}.

We showed in \citet{gg10} that this method of analysis introduces a weak bias that must be corrected for. We have followed the same procedure as in that study and use the comparison stars in the present work to correct for this bias. The bias correction was calculated as follows. We selected every other star from the comparison stars sample and treated them as if they were SWPs (''fake SWPs''), and we treated the remaining stars as comparison stars. We then calculated the weighted Li abundance differences as before (Figure 4a). Next, we exchanged the roles of the stars and repeated the analysis (Figure 4b). 
The best-fit lines are shown in Figures 4a and 4b. The slopes of the fits are $1.0 \times 10^{-3}$ and $6.0 \times 10^{-4}$dex K$^{-1}$, respectively; the average is close the value determined for the equivalent figure (Figure 7) in \citet{gg10}. We adopted an average least-squares linear fit to each figure and subtracted the fit from the data in Figure 3. The resulting bias-corrected data are shown in Figure 5.

\begin{figure}
  \includegraphics[width=3in]{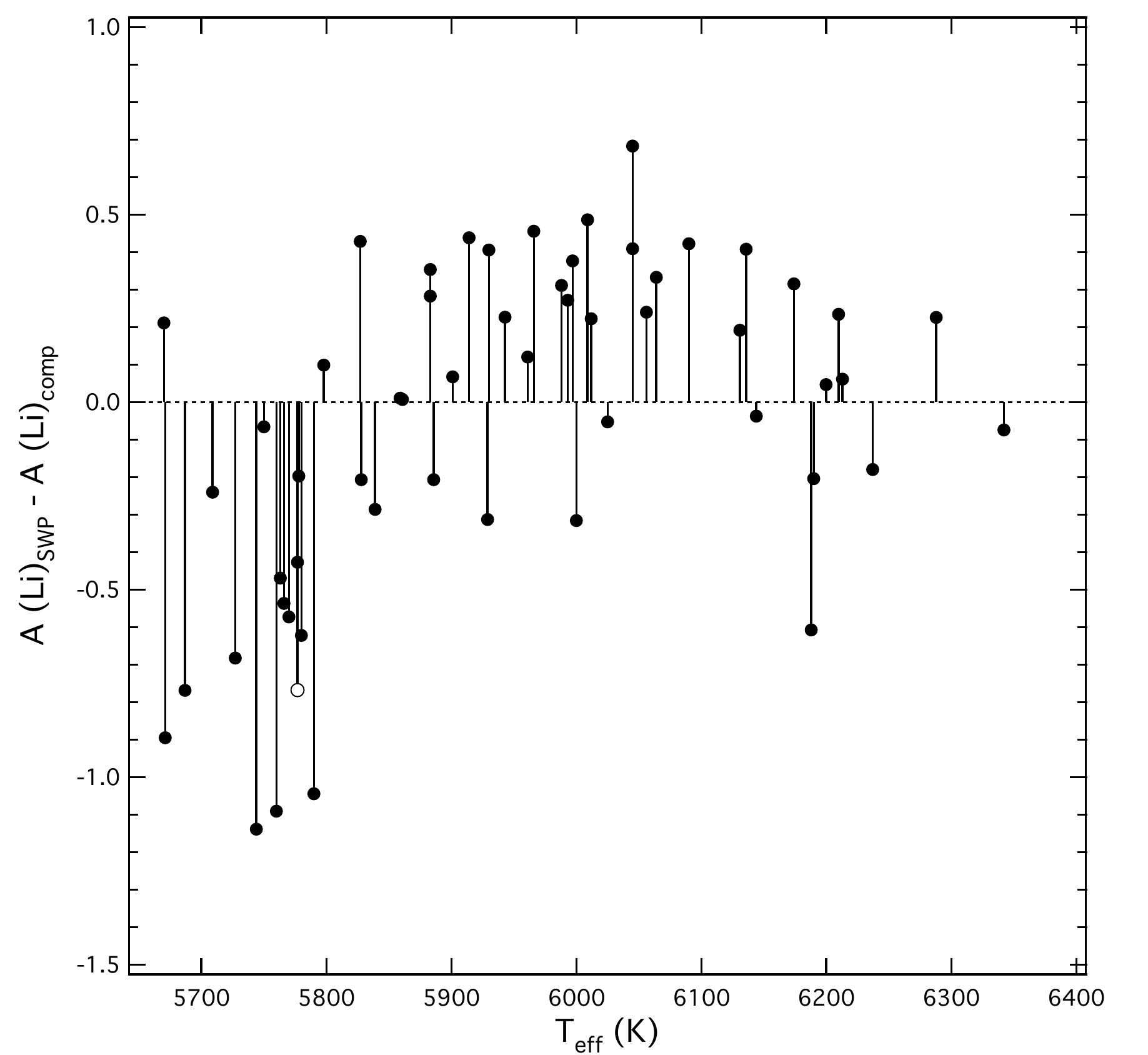}
 \caption{Weighted average Li abundance differences between SWPs and comparison stars. The open circle represents the Sun.}
\end{figure}

\begin{figure}
  \includegraphics[width=3in]{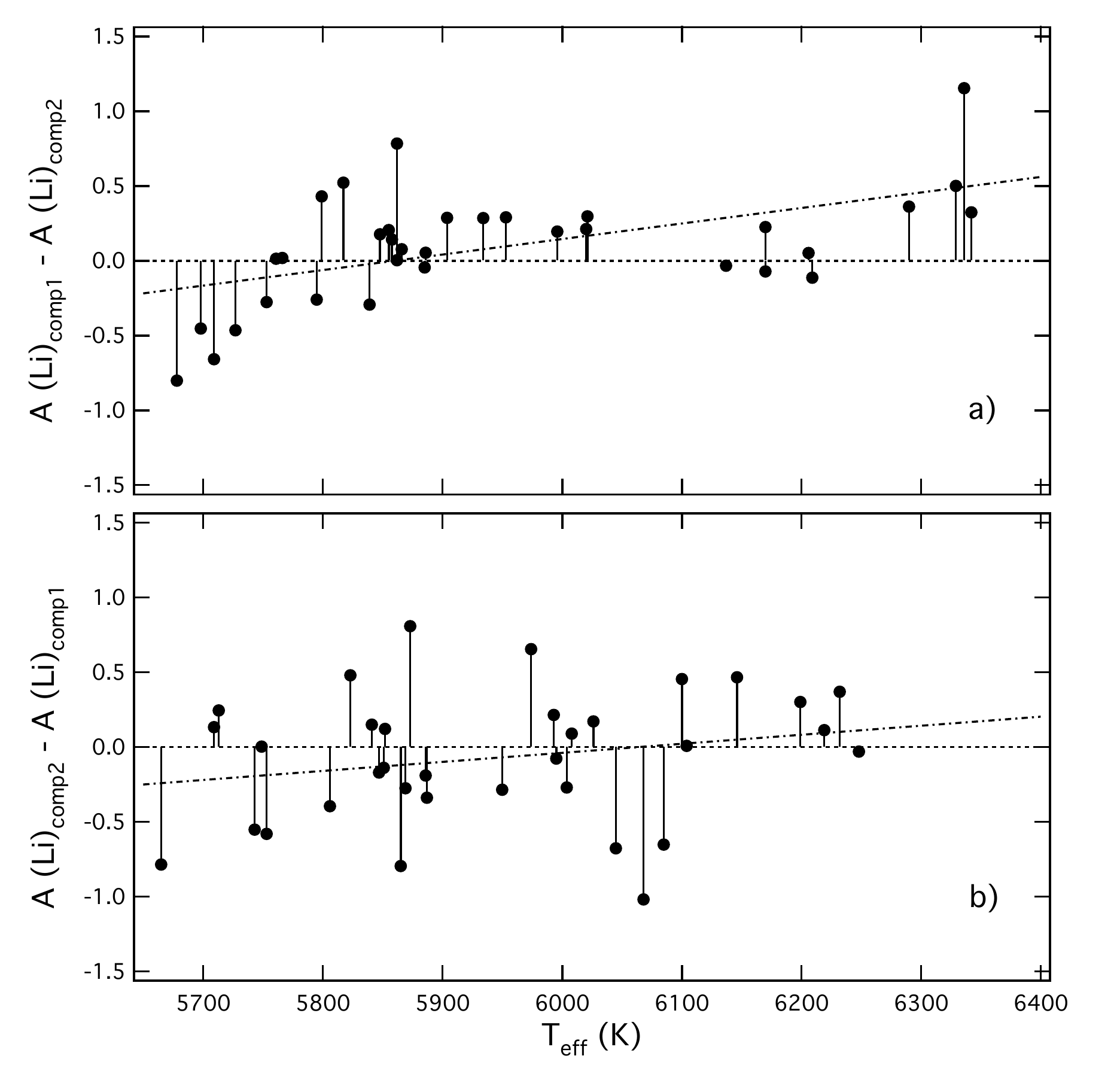}
 \caption{Weighted average Li abundance differences among the comparison stars. The roles of the comparison stars have been exchanged panel b compared to panel a. The least-squares fits are shown as dash-dotted lines. See text for details.}
\end{figure}

\begin{figure}
  \includegraphics[width=3in]{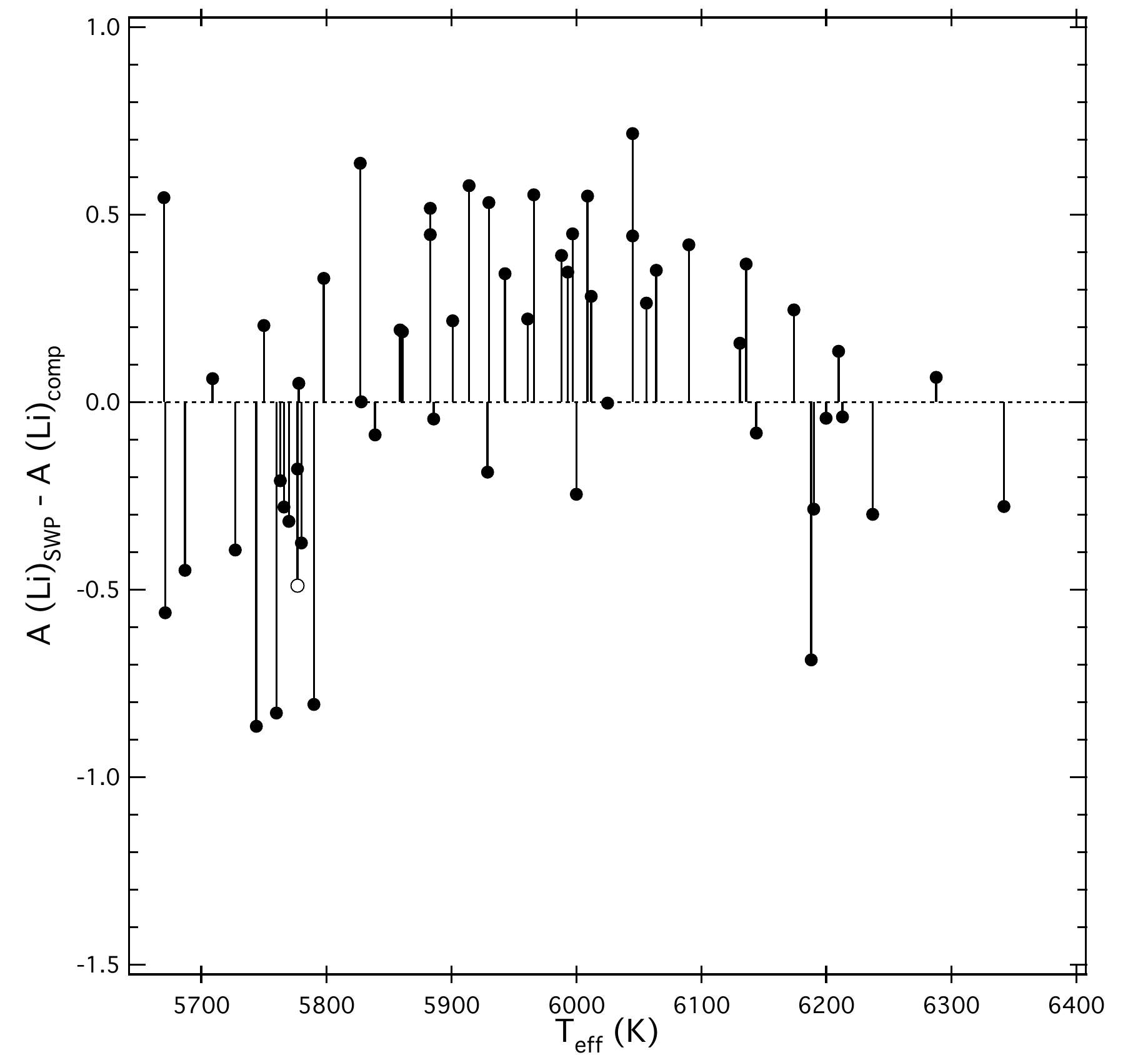}
 \caption{Same data as shown in Figure 3 but corrected for bias using average trend determined from data in Figure 4.}
\end{figure}

In Figure 6 we show the result of calculating the bias corrections using a different approach. In this case, we again made use of the comparison stars sample, but this time calculated a weighted-average Li abundance for each comparison star relative to all the other comparison stars. The slope of the least-squares fit to the data is $9.0 \times 10^{-4}$dex K$^{-1}$, which is very close to the average slope in Figure 4. However, a linear least-squares fit is not a good description of the data in this case. For this reason, we also calculated the average of the Li abundance differences in 100 K-wide bins, which we show as diamonds in the figure. We applied these binned corrections to the uncorrected data from Figure 3 to produce the bias-corrected SWP Li abundance weighted-differences in Figure 7.

\begin{figure}
  \includegraphics[width=3in]{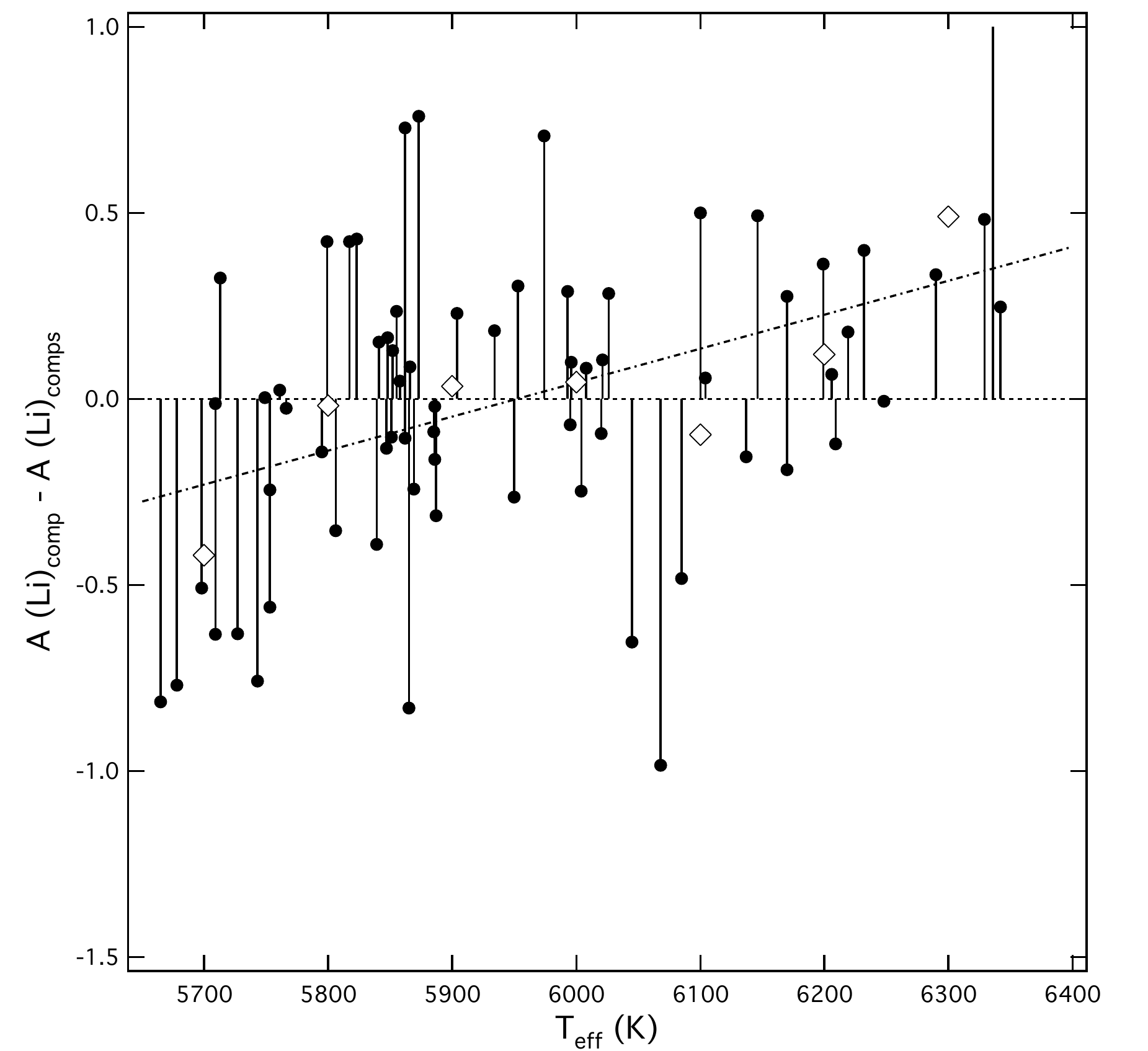}
 \caption{Weighted average Li abundance differences among the comparison stars. In this case, each comparison star has been compared to all the other comparison stars in our sample. The dashed-dotted line is a linear least-squares fit. The diamonds are averages in 100 K-wide bins. One point near the upper temperature limit is slightly off-scale.}
\end{figure}

\begin{figure}
  \includegraphics[width=3in]{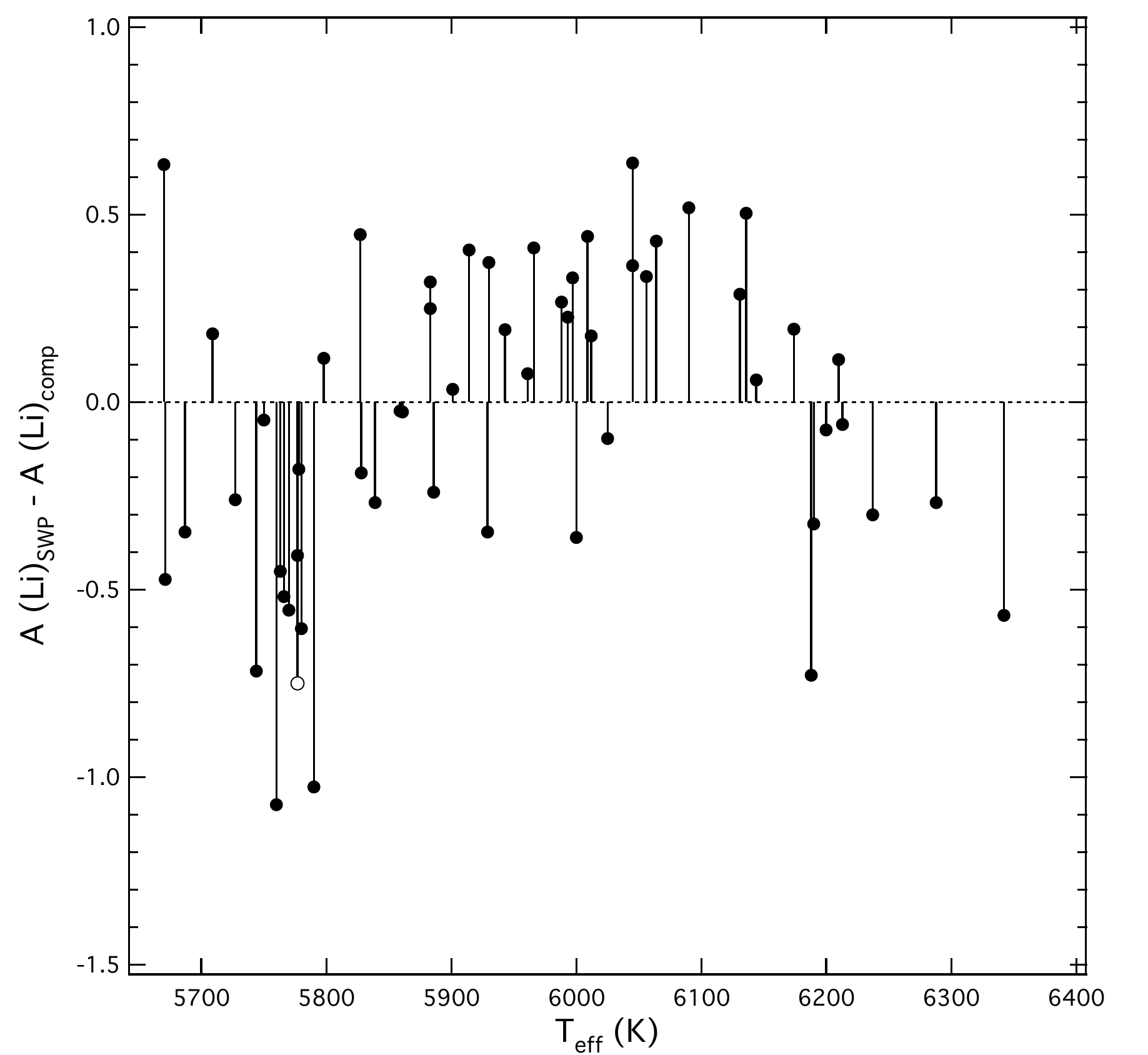}
 \caption{Data from Figure 3 corrected for bias using the binned corrections from Figure 6.}
\end{figure}

The weighting scheme used for our comparison is somewhat arbitrary, but it has the advantage that it is based on pure observables and makes the fewest assumptions. It is known, however, that a star's Li abundance declines with age. The age of a star determined from comparison of observations to theory depends on the details of the evolutionary model used, which is based on a number of assumptions. For example, it assumes that a star started its life with a homogeneous composition. This might not be the case with all SWPs (perhaps their convective envelopes were polluted by accreted planetary material). Nevertheless, we will apply the following age-based weighting scheme as a different approach to comparing Li abundances between SWPs and comparison stars. We begin by defining a new $\Delta$ index:
\begin{eqnarray*}
\Delta_{p,c}=30~\vert \log~{\rm T}_{\rm eff}^{c} - \log~{\rm T}_{\rm eff}^{p} \vert +
\vert {[Fe/H]}^{c} - {[Fe/H]}^{p} \vert \\
 + 0.5~\vert \log {g}^{\rm c} - \log {g}^{\rm p} \vert +\vert \log {\rm Age}^{c} - 
 \log {\rm Age}^{p} \vert
\end{eqnarray*}
where 'p'  refers to a SWP and  'c' refers to a comparison star. Two stars with identical values of T$_{\rm eff}$, $\log g$, [Fe/H] and age will have a $\Delta$ value of zero. This definition of the $\Delta$ index differs from our previous one only in the substitution of $\log$ age for M$_{\rm V}$. To calculate the weighted average difference between the Li abundance of a SWP and a set of comparison stars, we used the following equation (as before):
\begin{eqnarray*}
\Delta_{\rm Li,p}=\frac{\sum_{c=1}^{N}(\log Li_{p} - \log Li_{c})(\Delta_{p,c})^{-2}}{\sum_{c=1}^{N}(\Delta_{p,c})^{-2}}
\end{eqnarray*}
where the sums are taken over the number of comparison stars, N. We applied this alternative weighting scheme to our data, and we show the resulting weighted Li abundance differences ($\Delta_{\rm Li,p}$) in Figure 8. It is qualitatively similar in appearance to Figure 3.

We also performed an experiment with a $\Delta$ index that includes all the terms of the new $\Delta$ index above as well as the original M$_{\rm V}$ term. Figure 9 shows the resulting weighted Li abundance differences. It is very similar to Figure 8. This is not surprising, as the information about M$_{\rm V}$ is implicitly contained within the age parameter. Therefore, for the following analysis we will use only the $\Delta$ index defined by the equation above.

\begin{figure}
  \includegraphics[width=3in]{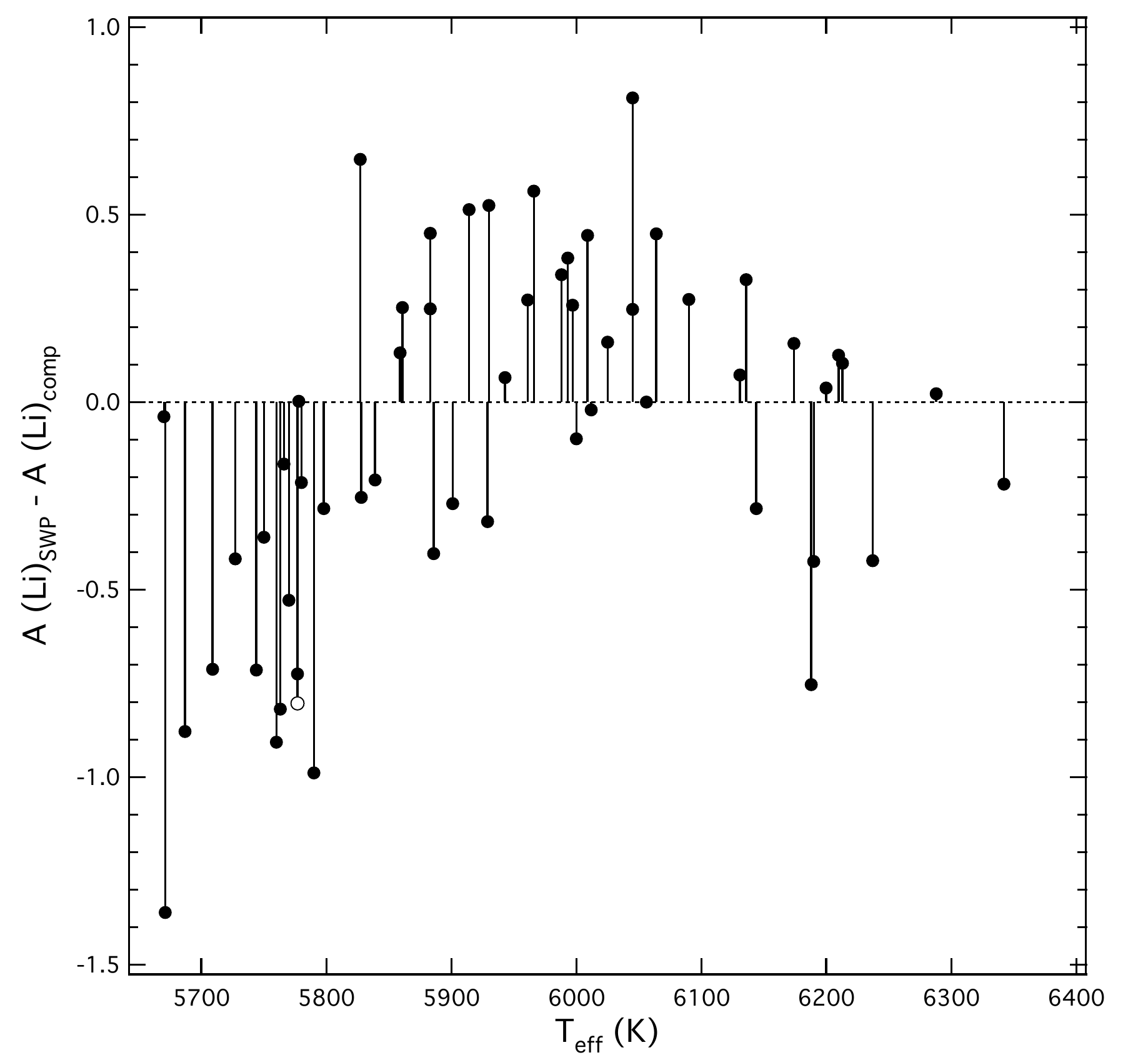}
 \caption{Same data as used in Figure 3 but calculated from the alternate weighting scheme (see text).}
\end{figure}

\begin{figure}
  \includegraphics[width=3in]{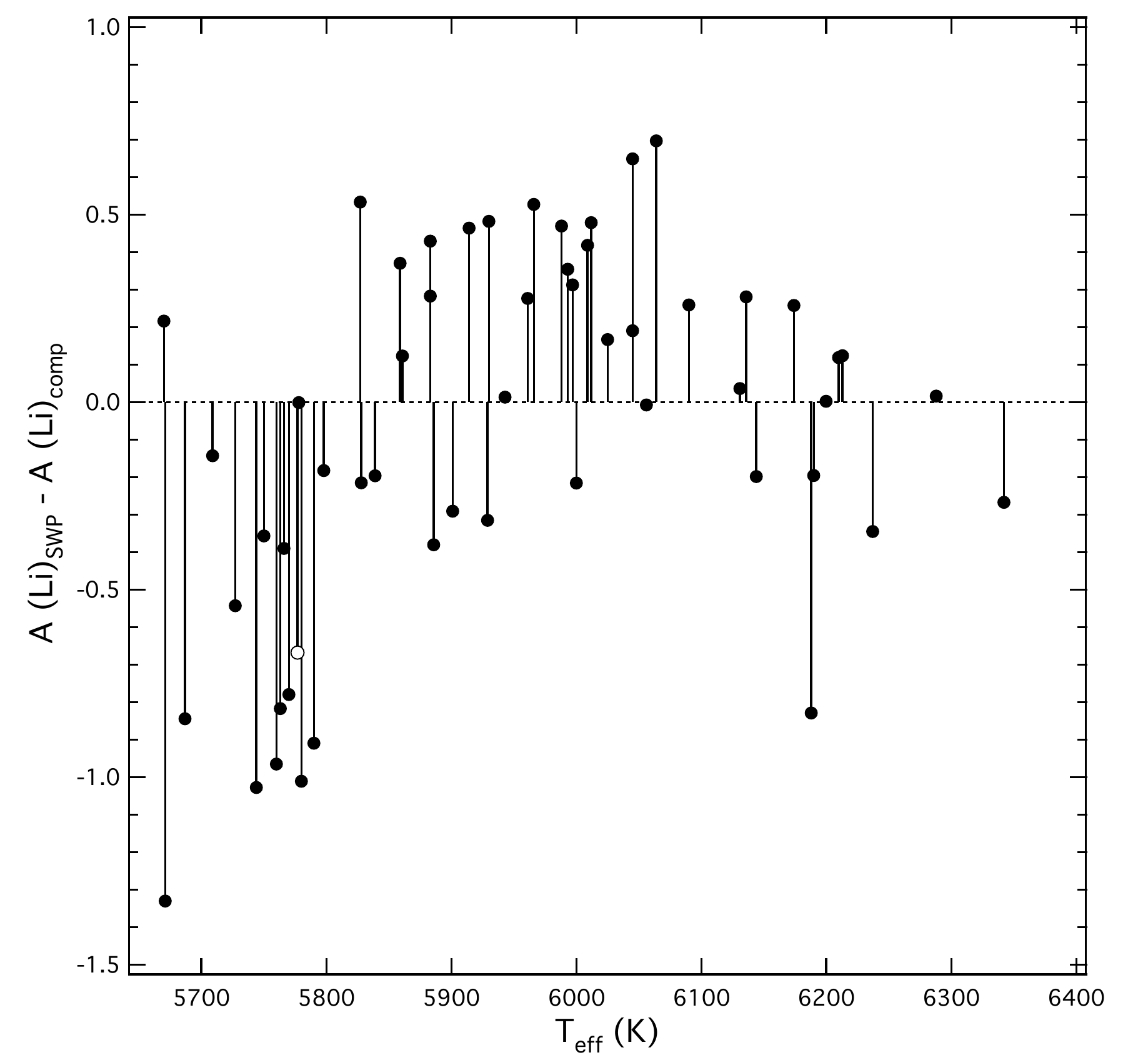}
 \caption{Same as Figure 8 but calculated from the alternate weighting scheme with five parameters.}
\end{figure}

\subsection{Adding literature data}

In order to test the robustness of the results shown in the previous section, we have expanded our dataset by adding data from the extensive compilation of \citet{ram12}; note, since their dataset already includes our data from \citet{gg10}, we only added the new data from the present study. We added our new data to their dataset with the following procedure, which were determined by comparing the stars in common to the two studies. First, our Li abundance values required a correction of +0.14 dex. No corrections were required for T$_{\rm eff}$, $\log g$, or [Fe/H]. We multiplied the age estimates for our stars by a factor of 1.2 to bring them into agreement with \citet{ram12}. When a star from our sample was already in theirs, we simply combined the star's parameters with weighted averages.

Next, we applied the following selection criteria to the combined dataset: stars must fall within the following ranges: $5500 <$ T$_{\rm eff} < 6400$, $\sigma$(T$_{\rm eff}) < 100$ K, [Fe/H] $> -0.70$, Li detected. Following this winnowing, the final sample size is 807 stars: 99 SWPs (or 100 with the Sun included), 241 non-SWPs, and 467 of unknown designation. We considered using this very large literature dataset in the following analysis, but the comparison stars likely include still undetected (but detectable) planets. For this reason, we also eliminated the 467 stars of unknown designation from the comparison sample. We will refer to these 100 SWPs and 241 non-SWPs as the ``literature dataset'' in the following analysis.

We show the weighted average Li abundance differences plot using the literature sample stars in Figure 10. The overal pattern is similar to that in Figure 3. Figure 11 shows the Li abundance differences among the comparison stars in the literature dataset. As we did with the data in plotted in Figure 6, we calculated the average bias correction for each 100 K wide bin in T$_{\rm eff}$. These average values were used to correct the data in Figure 10. The resulting bias-corrected weighted Li abundance differences are shown in Figure 12. Also shown in Figure 12 are the average difference in each T$_{\rm eff}$ bin along with the standard deviation of the average. For the bins centered at 5650 and 5750 K, the averages are smaller than zero Li abundance difference by 5 $\sigma$ and 3.8 $\sigma$, respectively. The two highest T$_{\rm eff}$ bins are only a little over 1 $\sigma$ below zero. The other bins do not deviate significantly from zero.

\begin{figure}
  \includegraphics[width=3in]{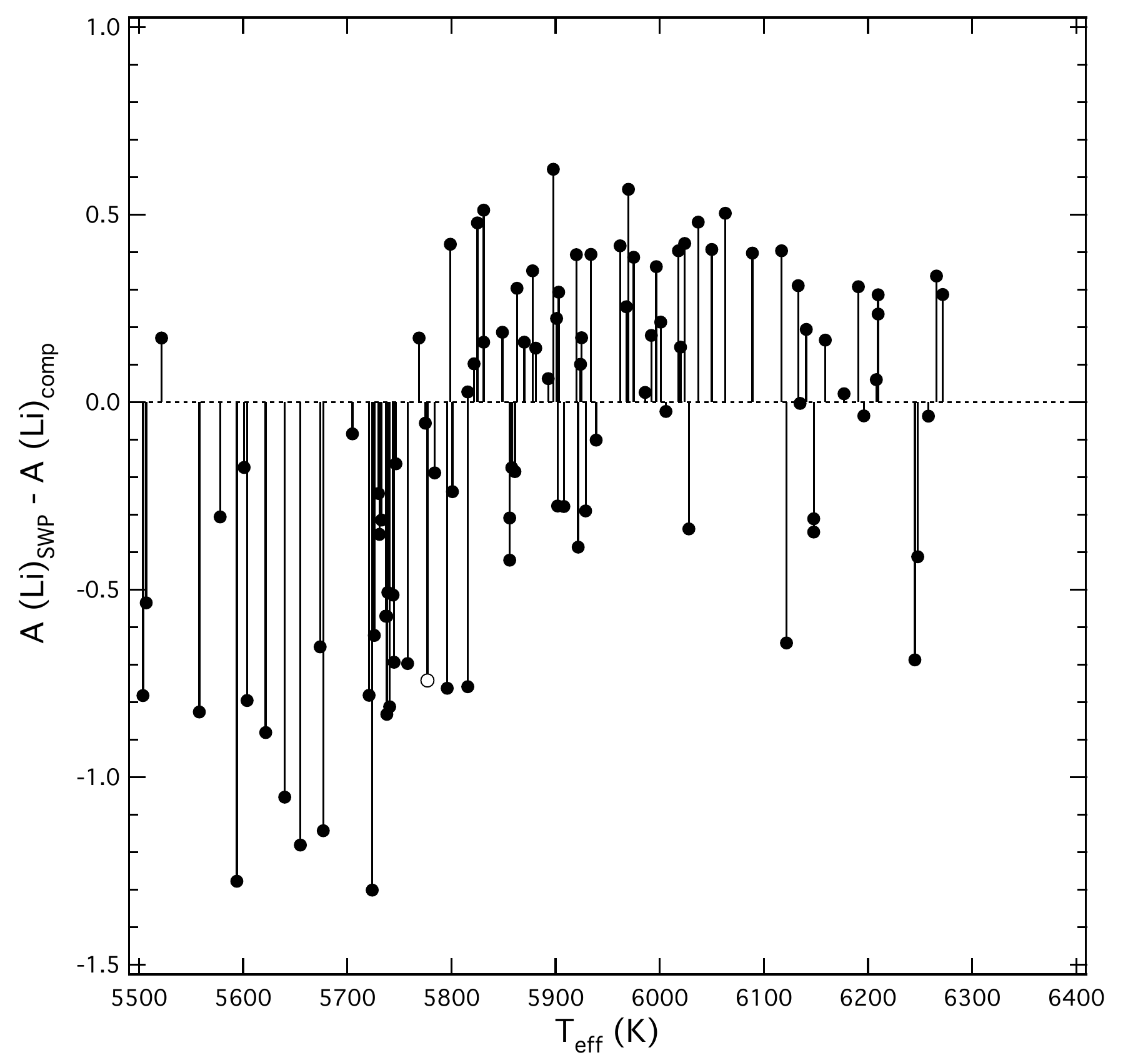}
 \caption{Weighted average Li abundance differences between SWPs and comparison stars from the literature dataset.}
\end{figure}

\begin{figure}
  \includegraphics[width=3in]{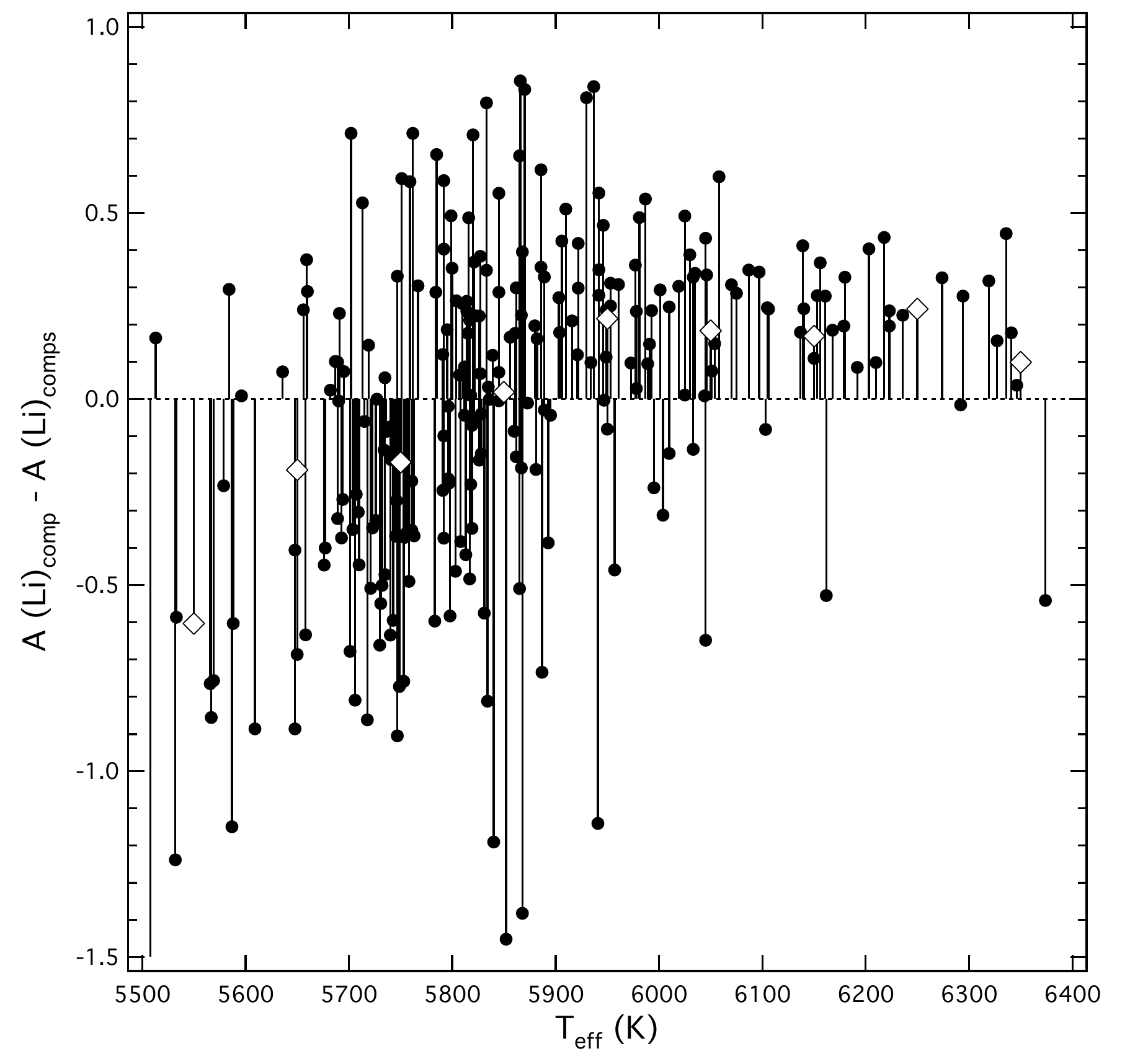}
 \caption{Weighted average Li abundance differences among the comparison stars from the literature dataset. The calculations were done in the same way as in Figure 6. The open diamonds are the averages of the Li abundance differences in 100 K wide bins.}
\end{figure}

\begin{figure}
  \includegraphics[width=3in]{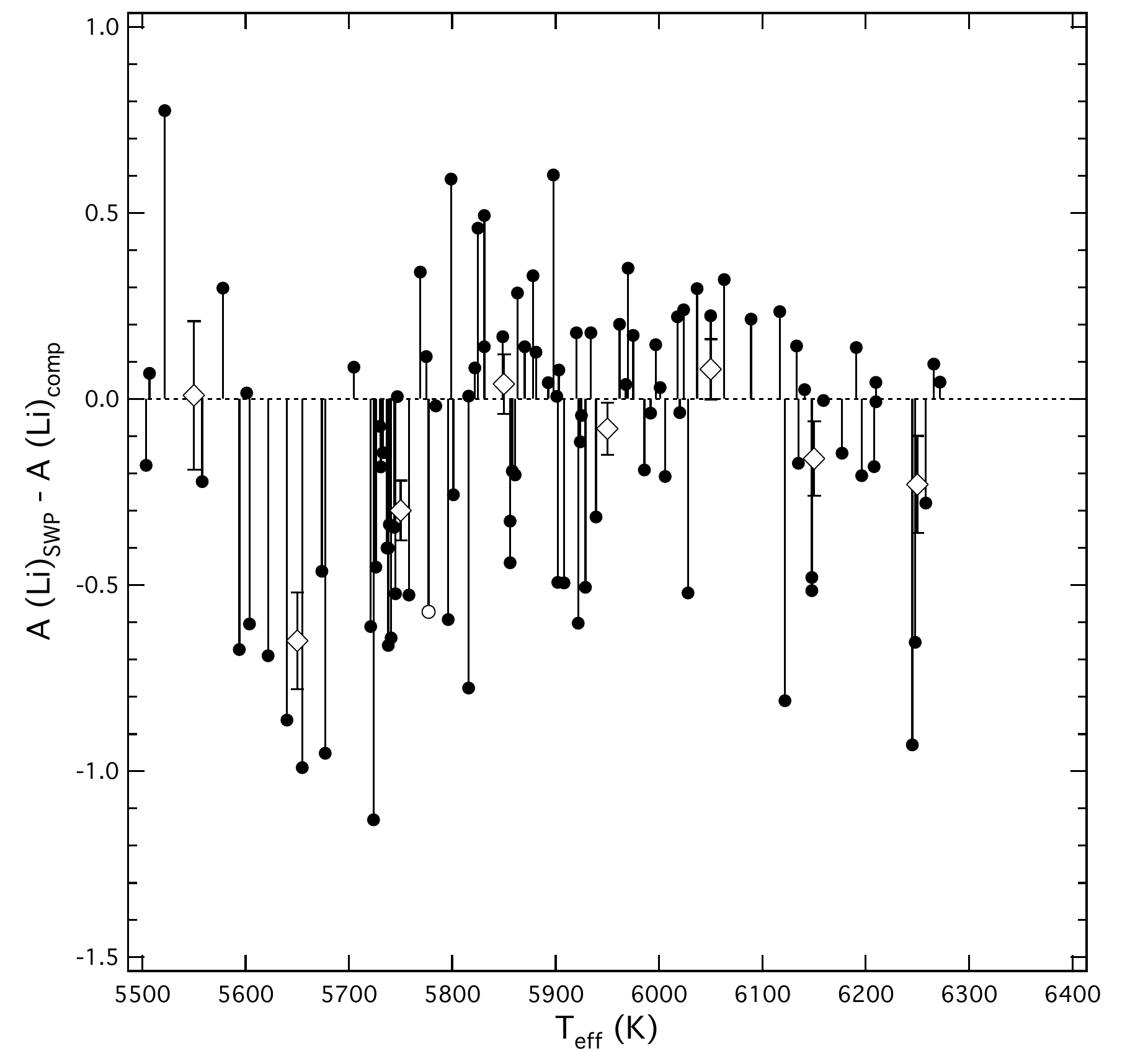}
 \caption{Data from Figure 10 corrected for bias using the binned bias corrections from Figure 11. The open diamonds represent the averages of the Li abundance differences in 100 K-wide bins, and the error bars correspond to the standard deviation of the average.}
\end{figure}

Perhaps the simplest way to compare the Li abundances in the SWP and non-SWP samples is simply to find the smallest difference in the $\Delta_{p,c}$ index between each SWP and a non-SWP comparison star; the Li abundance difference is then calculated between the SWP and the non-SWP star with the smallest $\Delta_{p,c}$ value. We have done this with the same dataset used to prepare Figures 10-12. The results are shown in Figure 13. No bias corrections were applied, as it is unlikely that it should be needed in this case. Unsurprisingly, the range of Li abundance differences is larger compared to the data in Figures 10 or 12. However, the overall pattern of the average Li abundances differences is very similar in Figures 12 and 13.

\begin{figure}
  \includegraphics[width=3in]{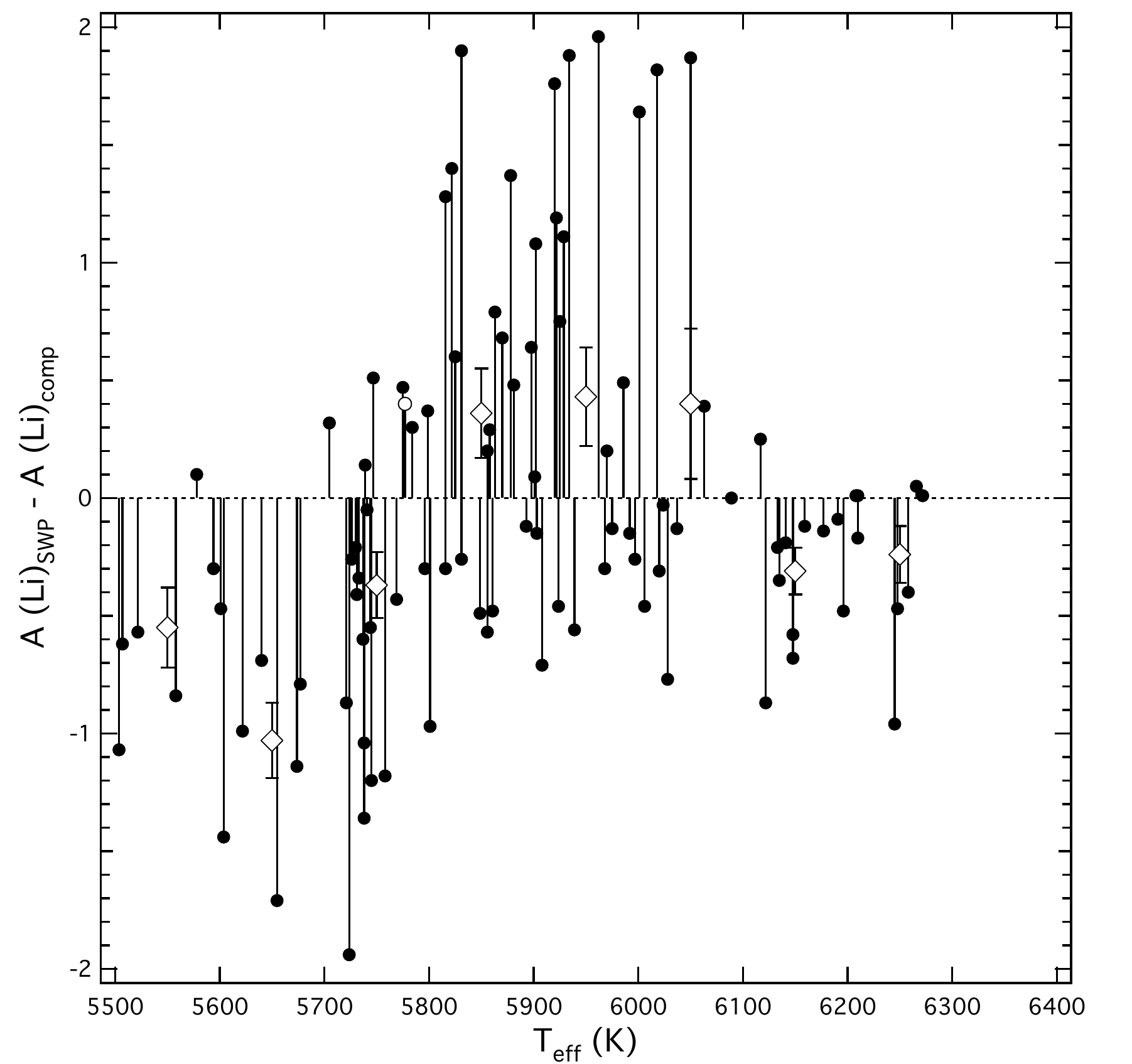}
 \caption{Li abundance difference between each SWP and the most similar comparison star. Symbols have same meanings as in Figure 12.}
\end{figure}

In Figures 12 and 13 the scatter in Li abundance differences at a given T$_{\rm eff}$ is much larger than the uncertainties in the individual measurements. No doubt some of the scatter is due to the lack of sufficiently similar SWPs and non-SWPs. For example, a given SWP and closest-matching non-SWP pair might have a similar set of measured T$_{\rm eff}$, $\log g$, and [Fe/H] values but differ significantly in the derived ages. This highlights the importance of including a large number of comparison stars in the analysis in order to properly sample the four parameters that are being compared.

\section{Conclusions}

We present the results of our analysis of high quality spectra of 37 late-F to early-G dwarfs, observed for the purpose of measuring Li in them. When combined with a large homogeneous sample of similar stars from the literature, we are able to confirm our previous findings from \citet{gg10} that the Li abundances of SWPs with T$_{\rm eff} \sim 5700$ K are smaller than those of stars without detected planets. In particular, SWPs with $5600 <$ T$_{\rm eff}  < 5800$ K are deficient in Li by about 0.5 dex relative to comparison stars with similar properties. There is weaker evidence that SWPs with T$_{\rm eff} > 6100$ K are also deficient in Li. Our results generally confirm other recent independent studies of Li abundances in SWPs \citep{tk10,del14}.

Additional observations of SWPs are needed for T$_{\rm eff} < 5600$ K and T$_{\rm eff} > 6100$ K to test whether Li abundance deviates significantly compared to non-SWPs in these regions. Observations of additional stars known not to have planets are required over the full temperature range studied here.

Work is also required on theoretical predictions of the amount of Li depletion expected from the interactions between a protoplanetary disk and its central star (e.g., \citet{bou08}). Any successful model will need to be able to explain the observed pattern of Li abundance differences between SWPs and non-SWPs with T$_{\rm eff}$ as well as the scatter.

\section*{Acknowledgments}

We thank Kyle A. McCarthy, a graduate student at the Department of Physics and Astronomy at the University of Kentucky, for obtaining spectra of HD 52711 for us. We also thank the anonymous reviewer for very helpful comments. This research has made use of the SIMBAD database, operated at CDS, Strasbourg, France.

\bsp

\label{lastpage}

\end{document}